\documentclass[twocolumn,secnumarabic,showpacs,preprintnumbers,nobibnotes,showfootnotes,amssymb,aps,prl]{revtex4}

\usepackage{graphicx}
\usepackage{dcolumn}
\usepackage{bm}
\usepackage{epsfig}
\newcommand{\newc}{\newcommand}

\def\beq{\begin{equation}}
\def\eeq{\end{equation}}
\def\bea{\begin{eqnarray}}
\def\eea{\end{eqnarray}}

\newc{\ie}{{\it i.e.~}}          \newc{\etal}{{\it et al.~}}
\newc{\eg}{{\it e.g.~}}          \newc{\etc}{{\it etc.~}}
\newc{\cf}{{\it c.f.~}}
\newc{\gsim}{\lower.7ex\hbox{$\;\stackrel{\textstyle>}{\sim}\;$}}
\newc{\lsim}{\lower.7ex\hbox{$\;\stackrel{\textstyle<}{\sim}\;$}}
\newc{\gev}{\,{\rm GeV}}
\newc{\mev}{\,{\rm MeV}}
\newc{\ev}{\,{\rm eV}}
\newc{\kev}{\,{\rm keV}}
\newc{\tev}{\,{\rm TeV}}

\newc{\gst}{\hat{g}_3}
\newc{\gw}{\hat{g}_2}
\newc{\gy}{\hat{g}_Y}
\newc{\g}{\hat{g}}
\newc{\muh}{\hat{\mu}}
\newc{\lambdah}{\hat{\lambda}}

\begin{document}

\title{Constraints on split-UED from Electroweak Precision Tests}

\author{Thomas Flacke}

\author{Christian Pasold}

\affiliation{Institut f\"ur Theoretische Physik und Astrophysik, 
Universit\"at  W\"urzburg, D-97074 W\"urzburg, Germany}

\begin{abstract}
We present strongly improved electroweak precision constraints on the split-UED model. We find that the dominating effect arises from contributions to the muon decay rate by the exchange of even-numbered $W$ boson Kaluza-Klein modes at tree-level, which so far have not been discussed in the context of UED models. The constraints on the split-UED parameter space are translated into bounds on the mass difference of the first Kaluza-Klein mode of fermions and the lightest Kaluza-Klein mode, which will be tested is the LHC. 
\end{abstract}

\date{\today}

\maketitle



In models with one universal extra dimension (UED)~\cite{Appelquist:2000nn}, all of the Standard Model particles are promoted to five-dimensional fields, propagating in a flat extra dimension (for earlier ideas closely related to UED models see Ref.~\cite{preUED}). In order to allow for the existence of chiral fermions at the zero-mode level, the extra dimension is chosen as an $S^1/\mathbb{Z}_2$ orbifold. We parameterize its fundamental domain as $x^5\equiv y\in [-\pi R/2,\pi R/2]$. The setup is symmetric under reflection at $y=0$, which implies a conserved $Z_2$ parity, so-called Kaluza-Klein (KK) parity.  One consequence of KK parity is that first KK mode particles can only be produced or annihilated in pairs. Among other consequences, this leads to a relatively weak bound on the KK mass scale $M_{KK}\equiv1/R$ of only  $M_{KK}\gsim 600 \gev$  from flavor constraints \cite{flavorbounds}, and  $M_{KK}\gsim 750 \, (300) \, \gev$ for $m_h= 115\, (750) \,\gev$ \cite{ewAY,ewGM,ewGfitter} from electroweak precision tests. Studies of the discovery reach of the first year LHC data suggest a bound of  $M_{KK}\gsim 700 \gev$ \cite{collbounds}.  Furthermore, KK parity guarantees the stability of the lightest KK particle (LKP), thus providing us with a potentially viable dark matter candidate \cite{servanttait}. For a review of the UED model and its phenomenology, see Ref.~\cite{Hooper:2007qk}.

The dark matter and the collider phenomenology of the UED model strongly depends on the detailed KK mass spectrum. The KK mass spectrum is modified when either  operators at the orbifold fixed points \cite{BLKTrefs} (in so-called ``non-minimal UED'') or 5D fermion mass terms  \cite{sUED1} (in so-called ``split-UED'') are taken into account. In both scenarios, not only the KK mass spectrum but also the couplings between KK particles are modified. In particular, the couplings of zero-mode fermions to even-numbered KK mode gauge bosons  are generically non-zero, even at tree-level.\footnote{In minimal UED - \ie in a UED setup without boundary localized operators or fermion mass terms - such KK number violating interactions are only induced at one-loop level \cite{CMS}.} Such couplings imply s-channel production and decay of the second KK mode gauge bosons at the LHC, which would result in  $Z'$-, $W'$-, and coloron-like signatures in di-lepton and di-jet channels \cite{2ndKKrefs,sUED2,sUEDWp}.

In this letter, we discuss the implications of  the altered mass spectrum and couplings of split-UED for constraints from electroweak precision tests.\footnote{The analysis for non-minimal UED will be presented in Ref.~\cite{WIP1}.} In the split-UED model,  five-dimensional fermion mass terms  
\beq
S_5\supset\int d^4x\int_{-\pi R/2}^{\pi R/2} -M_\Psi \overline{\Psi}\Psi
\eeq
are introduced for all fermions $\Psi=(Q,U,D,L,E)$. In order to preserve KK parity, the mass terms are chosen to have a KK parity-odd profile, which is taken to be proportional to  the Heavyside step function  \cite{sUED1}. For simplicity, we assume a universal 5D mass for all fermions $-M_Q=M_U=M_D=-M_L=M_E=\mu\,\theta(y)$.\footnote{These mass terms are universal in the sense that they lead to equal masses of all fermions at the $n^{th}$ KK level before spontaneous symmetry breaking. {\it C.f.} \cite{sUED2} for details.} 

Let us summarize some results from Ref.\cite{sUED2}, which are relevant for our analysis. The zero-mode fermions remain massless after KK decomposition. They only acquire a mass $m_f$ from spontaneous symmetry breaking via the Yukawa potential. The masses of the fermions at the first KK level follow from 
 \beq
 m^2_{f^{(1)}}= \mu^2 \pm k^2_1+m^2_{f} \mbox{\hspace{10pt}  for  \hspace{10pt} } \mu\gtrless\frac{2}{\pi R},
 \label{eq:mf}
 \eeq
 where $k_1$ is determined as the smallest positive solution of  \cite{sUED2}
 \bea
 \mu&=&-k_n \cot \left(\frac{k_n \pi R}{2}\right)\mbox{\hspace{13pt}  for  \hspace{8pt} } \mu>\frac{2}{\pi R}, \label{kquant}\\
 \mu&=&-k_1 \coth \left(\frac{k_1 \pi R}{2}\right)\mbox{\hspace{8pt}  for  \hspace{8pt} } \mu<\frac{2}{\pi R}.
 \eea
At all even-numbered KK levels one finds $m^2_{f^{(n)}} = \mu^2 +(n/R)^2+m^2_{f}$. At odd-numbered KK levels above the first level, the fermion masses squared are $m^2_{f^{(n)}} = \mu^2 +k_n^2+m^2_{f}$ with $k_n$  determined from the $\left(\frac{n+1}{2}\right)^{th}$ solution of Eq.~(\ref{kquant}), irrespective of whether $\mu$ is larger or smaller than $2/\pi R$.
The  zero-mode fermions couple to gauge bosons at even-numbered KK levels with a relative coupling strength of \cite{sUED2}
\beq
\frac{g_{002n}}{g_{000}}=\frac{(\mu R)^2 \left[1-\coth\left(\frac{\mu \pi R}{2}\right)\right]\left[1-(-1)^n e^{\mu \pi R}\right]}{\sqrt{2}\left[(\mu R)^2+n^2\right]},
\label{coupl}
\eeq
where $g_{000}$ is identified with the respective $SU(3)$, $SU(2)$, and $U(1)_Y$ Standard Model couplings. The couplings between zero-mode fermions and gauge fields at odd-numbered KK levels vanish due to KK parity. Let us also point out that no KK number violating couplings between zero-mode gauge bosons and KK mode fermions exist.\footnote{The zero-mode wave function for the gauge bosons is the same as in minimal UED and does not depend on the extra-dimensional coordinate $y$. Therefore, the orthogonality relations of the KK mode wave functions of the fermions guarantee the absence of KK number-violating couplings to KK fermions.}
 
After this summary of the KK masses and couplings, let us now turn to electroweak precision constraints. The constraints on the minimal UED model from electroweak precision tests have been studied in Refs.~\cite{ewAY,ewGM,ewGfitter} by calculating the UED contributions  $S_{UED},T_{UED}$ and $U_{UED}$ to the Peskin-Takeuchi parameters \cite{PT1} at one-loop-level. The leading contributions are \cite{ewAY,ewGM,ewGfitter}
\begin{widetext}
\bea
S_{UED}&=&\frac{4 \sin^2\theta_W}{\alpha}\left[\frac{3g^2}{4(4\pi)^2}\left(\frac{2}{9}\sum_n \frac{m^2_t}{(n/R)^2}\right)+\frac{g^2}{4(4\pi)^2}\left(\frac{1}{6}\frac{m^2_h}{1/R}\right)\zeta(2)\right], \label{Sloop}\\
T_{UED}&=&\frac{1}{\alpha}\left[\frac{3g^2}{2(4\pi)^2}\frac{m_t^2}{m_W^2}\left(\frac{2}{3}\sum_n \frac{m^2_t}{(n/R)^2}\right)+\frac{g^2\sin^2\theta_W}{(4\pi)^2\cos^2\theta_W}\left(-\frac{5}{12}\frac{m^2_h}{1/R}\right)\zeta(2)\right], \label{Tloop}\\
U_{UED}&=&-\frac{4\sin^2\theta_W}{\alpha}\left[\frac{g^2\sin^2\theta_W}{(4\pi)^2}\frac{m_W^2}{(1/R)^2}\left(\frac{1}{6}\zeta(2)-\frac{1}{15}\frac{m_h^2}{(1/R)^2}\zeta(4)\right)\right]. \label{Uloop}
\eea
\end{widetext}
The terms proportional to $\zeta(2)$ and  $\zeta(4)$ arise from sums over all KK Higgs loops and KK gauge loops. For the KK top loop contributions, we wrote out the explicit sum, because we will use the expressions above for split-UED, where the KK top masses are altered, such that the KK top loop contributions will not sum up to $\zeta(2)$. In minimal UED, the treatment in terms of oblique corrections is justified because vertex corrections are suppressed by $\alpha$ as compared to Eqs.(\ref{Sloop}-\ref{Uloop}) \cite{ewAY}. In split-UED, the same argument holds true, and the split-UED result is obtained by the simple replacement of $(n/R)^2$ with $k^2_n+\mu^2$ in the top-loop contributions in Eqs.~(\ref{Sloop},\ref{Tloop}).\footnote{Note that while KK parity is still conserved, KK number is violated in fermion-gauge-interactions. This leads to additional contributions in the one-loop corrections to propagators of non-zero KK mode gauge bosons. However, as argued earlier, no KK number violating interactions of zero-mode gauge bosons with KK fermions occur. As the $S_{UED}$, $T_{UED}$, and $U_{UED}$ parameters are determined from corrections to the zero-mode gauge boson propagators, no KK number violating processes contribute to them at one-loop level.} 

The above result holds for split-UED if the underlying electroweak parameters of the model are correctly identified in terms of the experimentally observed Fermi constant $G_f$, the fine-structure constant $\alpha$ at the $Z$-pole, and the $Z$ mass $m_Z$. For this matching, the couplings given in Eq.~(\ref{coupl}) play a central r\^ole. In split-UED, all even-numbered  $W$ KK modes contribute to the decay rate of the of the muon. The calculation of the Fermi constant yields
\beq
G_f=G^0_f+\delta G_f
\eeq
with
\beq
G^0_f= \frac{g^2}{\sqrt{32}  m_W^2} \mbox{ and } \delta G_f=\frac{1}{\sqrt{32}}\sum_n\frac{g^2_{002n}}{m_W^2+\left(\frac{2n}{R}\right)^2}.
\label{Gfcorr}
\eeq
$G^0_f$ is just the contribution from  the $W$ zero-mode, while  $\delta G_f$ denotes the sum of the contributions from all non-zero $W$ KK  modes.\footnote{In minimal UED, the same effect occurs, but as the couplings of zero-mode fermions to even-numbered KK mode gauge bosons are one-loop suppressed, contact interactions which are induced by an exchange of a non-zero gauge KK mode only occur at two-loop level.}  These contributions can be considered as corrections to a generalized gauge boson propagator, which includes the poles of all gauge boson KK modes. As we assumed equal 5D mass terms for all quarks and leptons, the couplings  given in Eq.~(\ref{coupl}) are universal, and a treatment in terms of oblique parameters is still viable - with one important modification: When performing a fit to the electroweak data at or near the $Z$-pole, the $W$ zero-mode is near its pole while all $W$ KK modes are off-resonance. Therefore, $S_{UED},T_{UED}$ and $U_{UED}$ depend only on $G^0_f$, and not on the experimentally measured value $G_f$, which differ by $\delta G_f$. As has been shown in Ref.\cite{CTWP} in the context of Randall-Sundrum models, such corrections to $G_f$ can be incorporated into the electroweak fit by matching the experimentally determined values of the new physics contributions $S_{NP},T_{NP}$ and $U_{NP}$ to effective parameters \footnote{For similar approaches see Refs. \cite{preCTWP}.}
\bea
S_{eff}&=&S_{UED},\nonumber\\
T_{eff}&=&T_{UED}-\frac{1}{\alpha}\frac{\delta G_f}{G_f} ,\nonumber\\
U_{eff}&=&U_{UED}+\frac{4 \sin^2\theta_W}{\alpha}\frac{\delta G_f}{G_f}. 
\label{STUeff}
\eea 
Using the expression for  $\delta G_f$ from Eq.~(\ref{Gfcorr}) together with Eqs.~(\ref{Sloop}-\ref{Uloop}) and inserting the masses and couplings determined from Eq.~(\ref{eq:mf}) and Eq.~(\ref{coupl}), the spilt-UED contributions to $S_{eff}$, $T_{eff}$ and $U_{eff}$ are implicitly given as functions of the three free parameters $\mu$, $R^{-1}$, and $m_h$. 

The experimental bounds on new-physics contributions to the oblique parameters have recently been updated by the \emph{Gfitter} collaboration \cite{ewGfitter}, who found 
\beq
S_{NP}=0.04\pm0.10\, ,\, T_{NP}=0.05\pm0.11\, , \, U_{NP}=0.08\pm0.11,
\label{STUexp}
\eeq
for a reference point $m_{h,ref}=120 \gev$ and $m_{t,ref}=173 \gev$ with  correlation coefficients  of $+0.89$ between $S_{NP}$ and $T_{NP}$, and $-0.45$  $(-0.69)$ between $S_{NP}$ and $U_{NP}$ ($T_{NP}$ and $U_{NP}$).

\begin{figure}[t]
\includegraphics[width=0.48\textwidth]{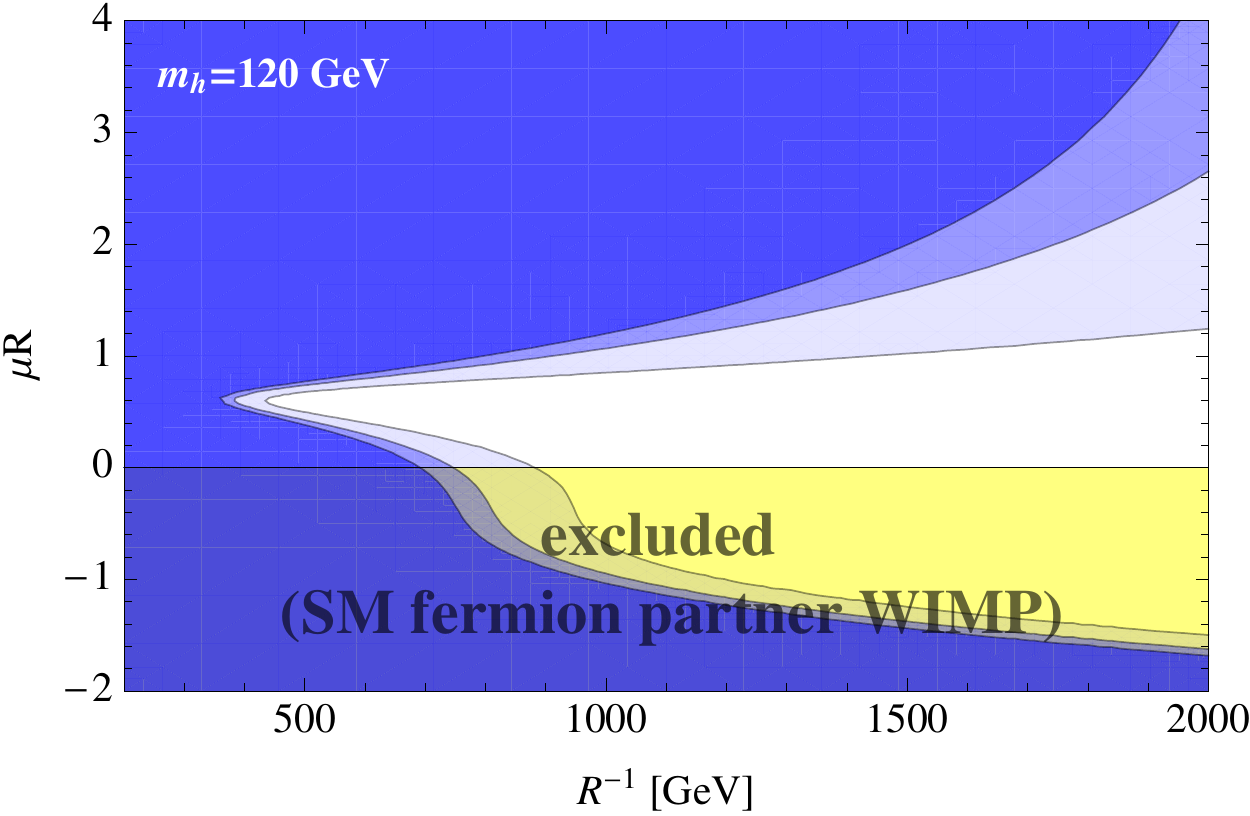}
\caption{Electroweak constraints on the split-UED parameter space $\mu R$ vs. $R^{-1}$. Shown are the  $65\%$, $95\%$ and $99\%$ c.l. fit contours. A 5D fermion mass parameter $\mu<0$ leads to a lightest Kaluza-Klein particle which is the KK partner of a Standard Model fermion and does not provide a viable dark matter candidate.}
\label{Fig:PS}
\end{figure}

\begin{figure}[t]
\includegraphics[width=0.48\textwidth]{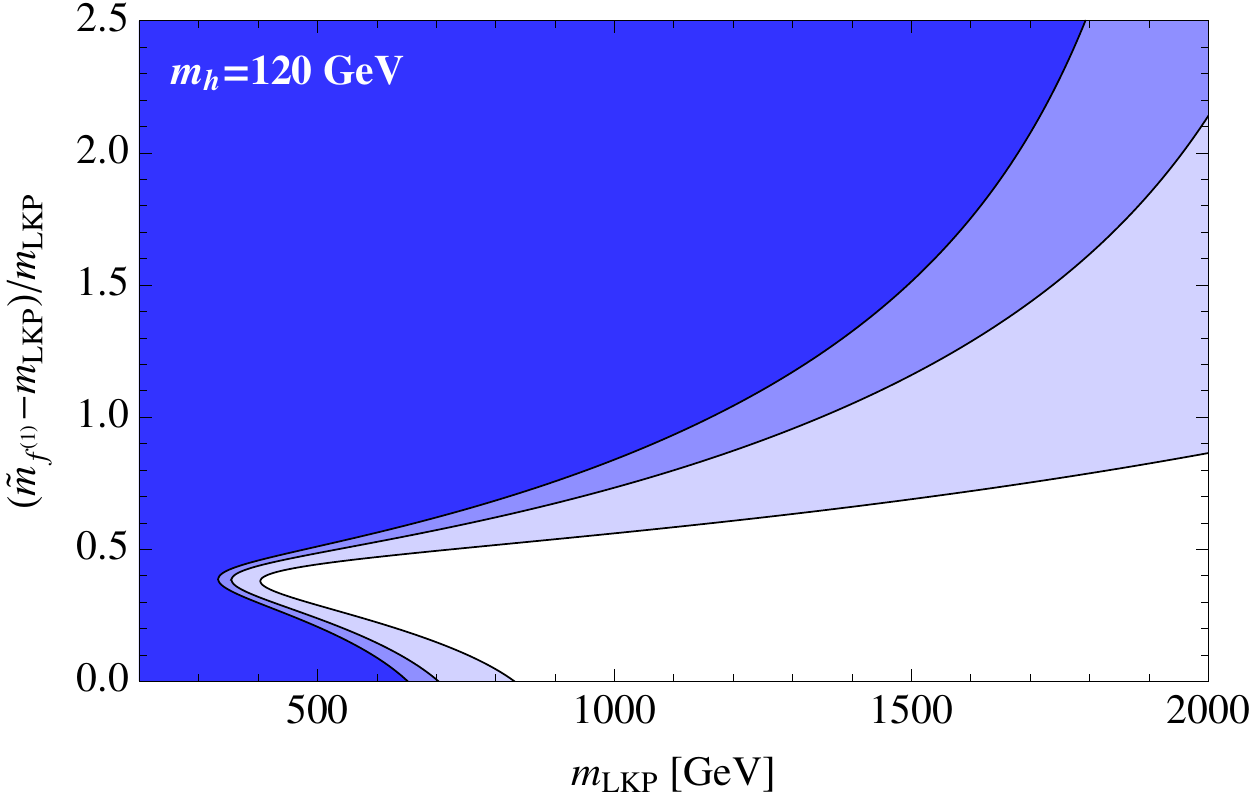}
\caption{Electroweak constraints on the mass splitting between the first KK mode fermions and the lightest KK particle at tree-level, as a function of the LKP mass. Shown are the $68\%$, $95\%$ and $99\%$ cl. fit contours which result from the bounds on the parameter space given in Fig. 1.}
\label{Fig:mdiff}
\end{figure}

We numerically calculate $S_{eff}$,  $T_{eff}$ and  $U_{eff}$ in the $\mu R$ vs. $R^{-1}$ parameter plane at the reference point $m_{h,ref}=120 \gev$ and $m_{t,ref}=173 \gev$. By performing a $\chi^2$-fit to the data given in Eq.~(\ref{STUexp}),  we obtain the 68\%, 95\% and 99\% confidence level contours in the split-UED parameter space, which are shown in Fig.~\ref{Fig:PS}.  For negative $\mu$, the KK neutrino becomes the lightest KK particle and thereby the dark matter candidate of the model. However, KK neutrino dark matter is excluded for $1/R\lesssim 50 \tev$ \cite{ST2}.\footnote{This bound could be avoided, if an additional dark matter candidate was added, into which the KK neutrino could decay (\cf \eg Ref.\cite{SUED3}). We therefore display the bounds also for negative $\mu$.}
The contours for $\mu\geq 0$ can be qualitatively understood as follows: For $m_h=120 \gev$, the Standard Model limit $R^{-1}\rightarrow \infty$ at $\mu=0$ is in agreement with the electroweak bounds. When decreasing $R^{-1}$, the UED loop contributions  $S_{UED}$ and $T_{UED}$ increase, whereby $T_{UED}$ yields the dominant contribution to $\chi^2$. This leads to a lower bound on $R^{-1}$ at $\mu=0$. For $\mu>0$,  $T_{eff}$ obtains a negative contribution which partially cancels the positive $T_{UED}$, thereby allowing for lower values of $R^{-1}$. However, for increasing $\mu$, $U_{eff}$ increases, and eventually exceeds the bound on $U_{NP}$. 

Using Eq.~(\ref{kquant}), we can translate the constraints from Fig.~\ref{Fig:PS} into bounds on the first KK mode fermion mass scale $\tilde{m}_{f^{(1)}}=\sqrt{\mu^2+k^2_1}$.  $\tilde{m}_{f^{(1)}}$ corresponds to the tree-level mass $m_{f^{(1)}}$ of Eq.~(\ref{eq:mf}) up to Yukawa contributions, which are negligible for all fermions apart from the top. Fig.~\ref{Fig:mdiff} shows the bounds on the relative mass splitting between the first KK mode fermions and the lightest KK particle $(\tilde{m}_{f^{(1)}}-m_{LKP})/m_{LKP}$ as a function of $R^{-1}$.  For the phenomenologically viable parameter region, the LKP is the first KK excitation of the $U(1)_Y$ gauge boson with $m_{LKP}=R^{-1}$. To use these bounds for comparison with physical masses, let us point out that loop contributions to the fermion KK masses can be sizable.  As has been shown in Ref.\cite{CMS} for minimal UED, the masses of the first KK mode quarks and leptons are raised by $~0.2/R$ and $0.05/R$ at one-loop-level, and similar results are to be expected in split-UED. This would imply that the constraint on the relative mass difference of the \emph{physical} KK fermion masses and the LKP at the first KK level is weaker by $\lesssim20\%$ ($\lesssim 5\%$) for KK quarks (fermions) than the values for $\tilde{m}_{f^{(1)}}$ shown in Fig.~\ref{Fig:mdiff}. A full one-loop analysis of the KK fermion masses is beyond the scope of this letter, so that the above is left as an estimate on the error of the bounds given in Fig.~\ref{Fig:mdiff}.

\emph{Summary and Outlook:}  We presented constraints on the split-UED parameter space with a universal 5D mass parameter $\mu$ from the electroweak precision observables $S,T$ and $U$. We found that corrections to the muon decay rate due to the exchange of even KK mode $W$ bosons lead to the dominant contributions to the effective $T$ and $U$ parameter. The resulting bounds on the split-UED parameter space shown in Fig.~\ref{Fig:PS} are substantially improved.  In particular, these constraints already exclude a substantial part of the split-UED parameter space currently tested in $Z'$  \cite{sUED2} and $W'$ searches \cite{sUEDWp} in the di-lepton channel at the LHC.

We translated this bound into a constraint on the relative difference between the masses of the first KK mode fermions and the LKP, which yields a prediction for a split-UED spectrum at LHC. We thereby showed, that even if 5D fermion mass terms are included, the UED spectrum must retain a certain degree of mass degeneracy, implying that the split-UED mass spectrum differs from typical SUSY spectra. 

For the constraints presented here, we assumed a universal 5D mass parameter for all fermions. If one allows for different mass parameters $\mu_L$ for leptons and $\mu_Q$ for quarks, the bound on $\mu_Q$ can be substantially weakened, because for $\mu_L=0$, the couplings of the muon to non-zero KK $W$ modes vanish, and muon decay only proceeds via the $W$ zero-mode.\footnote{For $\mu_L\neq\mu_Q$,  leptonic and hadronic channels at LEP are not affected universally anymore, so that a treatment in terms of the oblique $S,T$ and $U$ parameters is insufficient, and a global fit to the LEP data is required for a reliable electroweak analysis \cite{WIP2}.} However, $\mu_L=0$ also implies that the second KK modes of the electroweak gauge bosons cannot contribute to s-channel exchange in the di-lepton channel, which serves as one of the main search channels for UED and its extensions.

\bigskip

\emph{Acknowledgements: The authors would like to thank Carlos Wagner and Reinhold R\"uckl for valuable discussions. TF is supported in part by the Federal Ministry of Education and Research (BMBF) under contract number 05H09WWE.}


\bigskip

\begin{thebibliography}{99}
\bibitem{Appelquist:2000nn}
T.~Appelquist, H.~C.~Cheng and B.~A.~Dobrescu,
Phys.\ Rev.\ D {\bf 64} (2001) 035002
[arXiv:hep-ph/0012100].

\bibitem{preUED}
 I.~Antoniadis,
  Phys.\ Lett.\  B {\bf 246}, 377 (1990);
 I.~Antoniadis, N.~Arkani-Hamed, S.~Dimopoulos and G.~R.~Dvali,
  Phys.\ Lett.\  B {\bf 436}, 257 (1998)
  [arXiv:hep-ph/9804398].


\bibitem{flavorbounds}  
A.~J.~Buras, M.~Spranger and A.~Weiler,
Nucl.\ Phys.\ B {\bf 660} (2003) 225
[arXiv:hep-ph/0212143];
A.~J.~Buras, A.~Poschenrieder, M.~Spranger and A.~Weiler,
Nucl.\ Phys.\ B {\bf 678} (2004) 455
[arXiv:hep-ph/0306158];
  U.~Haisch and A.~Weiler,
  Phys.\ Rev.\  D {\bf 76}, 034014 (2007)
  [arXiv:hep-ph/0703064];
 P.~Colangelo, F.~De Fazio, R.~Ferrandes and T.~N.~Pham,
  Phys.\ Rev.\  D {\bf 73}, 115006 (2006)
  [arXiv:hep-ph/0604029];
  Phys.\ Rev.\  D {\bf 74}, 115006 (2006)
  [arXiv:hep-ph/0610044];
  T.~M.~Aliev and M.~Savci,
  Eur.\ Phys.\ J.\  C {\bf 50}, 91 (2007)
  [arXiv:hep-ph/0606225].

\bibitem{ewAY}
T.~Appelquist and H.~U.~Yee,
Phys.\ Rev.\ D {\bf 67} (2003) 055002 [arXiv:hep-ph/0211023];
\bibitem{ewGM}
  I.~Gogoladze and C.~Macesanu,
  Phys.\ Rev.\  D {\bf 74}, 093012 (2006)
  [arXiv:hep-ph/0605207];
\bibitem{ewGfitter}
 M.~Baak {\it et al.} [The G-fitter Group],
  arXiv:1107.0975 [hep-ph].

\bibitem{collbounds}
B.~Bhattacherjee and K.~Ghosh,
  Phys.\ Rev.\ D\ {\bf 83}, 034003  (2011)
  [arXiv:1006.3043 [hep-ph]],
 H.~Murayama, M.~Nojiri and K.~Tobioka,
  arXiv:1107.3369 [hep-ph].
  A.~Datta, A.~Datta and S.~Poddar,
  arXiv:1111.2912 [hep-ph].

\bibitem{servanttait}
G.~Servant and T.~M.~P.~Tait,
Nucl.\ Phys.\ B {\bf 650} (2003) 391
[arXiv:hep-ph/0206071].

\bibitem{Hooper:2007qk}
  D.~Hooper and S.~Profumo,
  Phys.\ Rept.\  {\bf 453}, 29 (2007)
  [arXiv:hep-ph/0701197].


\bibitem{BLKTrefs}
  G.~R.~Dvali, G.~Gabadadze, M.~Kolanovic and F.~Nitti,
  Phys.\ Rev.\  D {\bf 64}, 084004 (2001)
  [arXiv:hep-ph/0102216];
  M.~S.~Carena, T.~M.~P.~Tait and C.~E.~M.~Wagner,
  Acta Phys.\ Polon.\  B {\bf 33} (2002) 2355
  [arXiv:hep-ph/0207056];
  F.~del Aguila, M.~Perez-Victoria and J.~Santiago,
  JHEP {\bf 0302}, 051 (2003)
  [arXiv:hep-th/0302023];
  F.~del Aguila, M.~Perez-Victoria and J.~Santiago,
  JHEP {\bf 0610}, 056 (2006)
  [arXiv:hep-ph/0601222].
  T.~Flacke, A.~Menon and D.~J.~Phalen,
  Phys.\ Rev.\ D\ {\bf 79}, 056009  (2009)
  [arXiv:0811.1598 [hep-ph]].
  
  \bibitem{sUED1}
 S.~C.~Park and J.~Shu,
  Phys.\ Rev.\ D\ {\bf 79}, 091702  (2009)
  [arXiv:0901.0720 [hep-ph]].
  
  
  \bibitem{2ndKKrefs}
    A.~Datta, K.~Kong and K.~T.~Matchev,
  Phys.\ Rev.\ D\ {\bf 72}, 096006  (2005)
  [Erratum-ibid.\ D\ {\bf 72}, 119901  (2005)]
  [hep-ph/0509246],
  
  
   \bibitem{sUED2} 
   K.~Kong, S.~C.~Park and T.~G.~Rizzo,
  JHEP\ {\bf 1004}, 081  (2010)
  [arXiv:1002.0602 [hep-ph]].
  
   \bibitem{sUEDWp} 
  D.~Kim, Y.~Oh and S.~C.~Park,
  arXiv:1109.1870 [hep-ph].

 
\bibitem{PT1} 
  M.~E.~Peskin and T.~Takeuchi,
  Phys.\ Rev.\ Lett.\ \ {\bf 65}, 964  (1990).


\bibitem{CTWP}
  M.~S.~Carena, E.~Ponton, T.~M.~P.~Tait and C.~E.~MWagner,
  Phys.\ Rev.\ D\ {\bf 67}, 096006  (2003)
  [hep-ph/0212307].


\bibitem{ST2}
  G.~Servant and T.~M.~P.~Tait,
  New J.\ Phys.\ \ {\bf 4}, 99  (2002)
  [hep-ph/0209262].
  
\bibitem{CMS} 
  H.~-C.~Cheng, K.~T.~Matchev and M.~Schmaltz,
  Phys.\ Rev.\ D\ {\bf 66}, 036005  (2002)
  [hep-ph/0204342].

\bibitem{WIP1}
T.~Flacke, in preparation.

\bibitem{preCTWP}
  T.~G.~Rizzo and J.~D.~Wells,
  Phys.\ Rev.\ D\ {\bf 61}, 016007  (2000)
  [hep-ph/9906234],
  H.~Davoudiasl, J.~L.~Hewett and T.~G.~Rizzo,
  Phys.\ Lett.\ B\ {\bf 473}, 43  (2000)
  [hep-ph/9911262],
    C.~Csaki, J.~Erlich and J.~Terning,
  Phys.\ Rev.\ D\ {\bf 66}, 064021  (2002)
  [hep-ph/0203034].

\bibitem{SUED3}
  K.~Kong, S.~C.~Park and T.~G.~Rizzo,
  JHEP\ {\bf 1007}, 059  (2010)
  [arXiv:1004.4635 [hep-ph]].

\bibitem{WIP2}
T.~Flacke, in preparation.



\end{thebibliography}
\end{document}